\documentclass[conference]{IEEEtran}
\IEEEoverridecommandlockouts
\usepackage{cite}
\usepackage{amsmath,amssymb,amsfonts}
\usepackage{algorithm}
\usepackage{algorithmic}
\usepackage{graphicx}
\usepackage{graphics}
\usepackage{textcomp}
\usepackage{cancel}

\usepackage{subcaption}
\usepackage{multirow}
\usepackage[inline]{enumitem}
\usepackage[dvipsnames,svgnames]{xcolor}
\usepackage{gensymb}
\usepackage{booktabs}
\usepackage{adjustbox}

\def\BibTeX{{\rm B\kern-.05em{\sc i\kern-.025em b}\kern-.08em
    T\kern-.1667em\lower.7ex\hbox{E}\kern-.125emX}}

\usepackage[numbers,sort&compress,square]{natbib}
\usepackage[]{hyperref}

\usepackage[mathlines,switch]{lineno}
\makeatletter

\let\old@ps@IEEEtitlepagestyle\ps@IEEEtitlepagestyle
\def\confheader#1{%
    \def\ps@IEEEtitlepagestyle{%
        \old@ps@IEEEtitlepagestyle%
        \def\@oddhead{\strut\hfill#1\hfill\strut}%
        \def\@evenhead{\strut\hfill#1\hfill\strut}%
    }%
    \ps@headings%
}
\makeatother

\begin{document}

\title{Exploration of Unary Arithmetic-Based Matrix Multiply Units for Low Precision DL Accelerators}


\author{
    \IEEEauthorblockN{Prabhu Vellaisamy\IEEEauthorrefmark{1}, Harideep Nair\IEEEauthorrefmark{1}, Di Wu\IEEEauthorrefmark{2}, Shawn Blanton\IEEEauthorrefmark{1}, and John Paul Shen\IEEEauthorrefmark{1}}
    \IEEEauthorblockA{\IEEEauthorrefmark{1}Electrical and Computer Engineering Department, Carnegie Mellon University
    \\\{pvellais, hpnair, rblanton, jpshen\}@andrew.cmu.edu}
    \IEEEauthorblockA{\IEEEauthorrefmark{2}Department of Electrical and Computer Engineering, University of Central Florida
    \\di.wu@ucf.edu}
}


\maketitle

\begin{abstract}

General matrix multiplication (GEMM) is a fundamental operation in deep learning (DL). 
With DL moving increasingly toward low precision, recent works have proposed novel unary GEMM designs 
as an alternative to conventional binary GEMM hardware. A rigorous evaluation of recent unary and binary GEMM designs is needed to assess the potential of unary hardware for future DL compute. This paper focuses on unary GEMM designs for integer-based DL inference and performs a detailed evaluation of three latest unary design proposals, namely, uGEMM, \textit{tu}GEMM and \textit{tub}GEMM, by comparing them to a conventional binary GEMM.
Rigorous post-synthesis evaluations beyond prior works are performed across varying bit-widths and matrix sizes to assess the designs' tradeoffs and determine optimal sweetspots. Further, we perform weight sparsity analysis across eight pretrained convolutional neural networks (CNNs) and the LLaMA2 large language model (LLM). In this work we demonstrate how unary GEMM can be effectively used for energy-efficient compute in future edge AI accelerators. 

\end{abstract}


\section{Introduction}

Recent advancements in artificial intelligence (AI), particularly deep learning (DL), have led to notable achievements, surpassing human performance in tasks like image and speech recognition. Despite these successes, computation cost for running DL models is ever-increasing at an exponential rate \cite{openai}. This has led to the development of deep learning accelerators (DLAs) with dedicated hardware to optimize general matrix multiplication (GEMM), the fundamental operation in DL. Prominent examples include tensor cores in modern GPUs and matrix multiplication units (MXUs) in Google's Tensor Processing Units (TPUs). Additionally, devices such as edge TPU and NVIDIA's Jetson exemplify the trend toward edge computing. These DLAs leverage binary arithmetic optimizations and are characterized by their use of low-precision arithmetic and optimized dataflow to achieve hardware efficiency.

Unary computing, touted as a promising alternative to relatively complex binary arithmetic, offers unique advantages in computational efficiency, especially for low-precision tasks prevalent in AI/DL. It manifests in two forms: (i) \textit{rate-coding} and (ii) \textit{temporal-coding}.
In rate-unary computing, data is encoded based on the frequency of 1s, directly proportional to the represented value. Temporal-unary encoding represents data as a consecutive sequence of 1s followed by 0s, with the number of 1s indicating the data value.
Traditionally, rate-unary encoding allows for low-complexity arithmetic hardware suitable for stochastic computing. It significantly simplifies circuits, such as using single AND gates as multipliers and multiplexers as adders \cite{alaghi2017promise}, providing a reasonable approximation of the ideal result with an inherent accuracy compromise. In contrast, temporal-unary encoding enables exact deterministic compute.

Among the latest innovations in this field are three unary GEMM architectures: (i) uGEMM, a unified rate-and-temporal-encoded design executing stochastic GEMM operations \cite{ugemm_paper}, (ii) \textit{tu}GEMM, the first fully-temporal GEMM design ensuring deterministic compute with full accuracy \cite{tugemm}, and (iii) \textit{tub}GEMM, a novel temporal-binary hybrid successor to \textit{tu}GEMM that significantly reduces latency with relatively modest hardware overhead, improving overall energy efficiency \cite{vellaisamy2023tubgemm}. Despite comparisons against previous unary GEMM approaches, a holistic evaluation of the unary GEMM designs against traditional binary GEMM designs in today's DLAs has not been explored in literature.

Unary computing gains prominence in DL inference as the industry gravitates toward lower precision computing, propelled by advancements in quantization techniques. Although precision scaling poses challenges during training, strides have been made in retaining accuracy from FP32 (32-bit floating point) to as low as FP8 (8-bit floating point) for training and as low as INT4 and INT2 (4-bit and 2-bit integer formats, respectively) for inference \cite{wang20188, choi2019accurate}.
In the domain of large language models (LLMs), NVIDIA's Grace Hopper, with its FP8 Transformer Engine, demonstrates significant speedups for transformer-based models. Latest works like BitNet b1.58 \cite{ma2024era}, utilizing ternary weights, pave the way for 1-bit LLMs.

This paper aims to explore the potential of unary GEMM in enhancing the computational efficiency of edge DLAs for low-precision AI inference. By comparing three latest unary GEMM designs that outperform prior unary works with traditional binary GEMM, we seek to understand their relative strengths and limitations. This comparative analysis, complemented by sparsity profiling of DL models, aims to provide insights and recommendations for future DLAs to facilitate more sustainable edge AI deployment. 
%
Key contributions are:
\begin{itemize}
    \item Current landscape for unary GEMM compute is largely unevaluated, with unary-based designs developed in an \textit{ad-hoc} manner. Our work is a first attempt at contextualizing the latest unary research for INT-based AI inference by juxtaposing and evaluating three recent unary GEMM designs against a traditional binary GEMM. Prior works \cite{ugemm_paper}\cite{tugemm}\cite{vellaisamy2023tubgemm} only compare against other unary GEMM.

    \item Previous works \cite{ugemm_paper}\cite{tugemm}\cite{vellaisamy2023tubgemm} utilize different process technologies (TSMC45 vs. Nangate45) and compare designs using only one configuration (8-bit, 16x16 GEMM). This paper extends beyond those works by comparing across varying bit-widths and matrix sizes using a single technology (Nangate45). This enables fair comparison across all architectures, providing a more holistic evaluation.
    
    \item Weight sparsities of eight CNNs and LLaMA2 LLM are profiled to assess unary designs' ability to inherently exploit sparsity to improve latency and energy efficiency.
    
    
    \item Design tradeoffs are assessed amongst the unary and binary designs, and the optimal design sweetspot for hardware efficiency is determined. Further, potential future directions for edge DLAs are discussed, anticipating a new wave of unary arithmetic research for AI compute.
\end{itemize}


\section{Background on Unary GEMM Architectures}
\label{sec:gemms}

\begin{figure}[t]
\centering
\includegraphics[width=0.9\columnwidth, height=7.3cm]{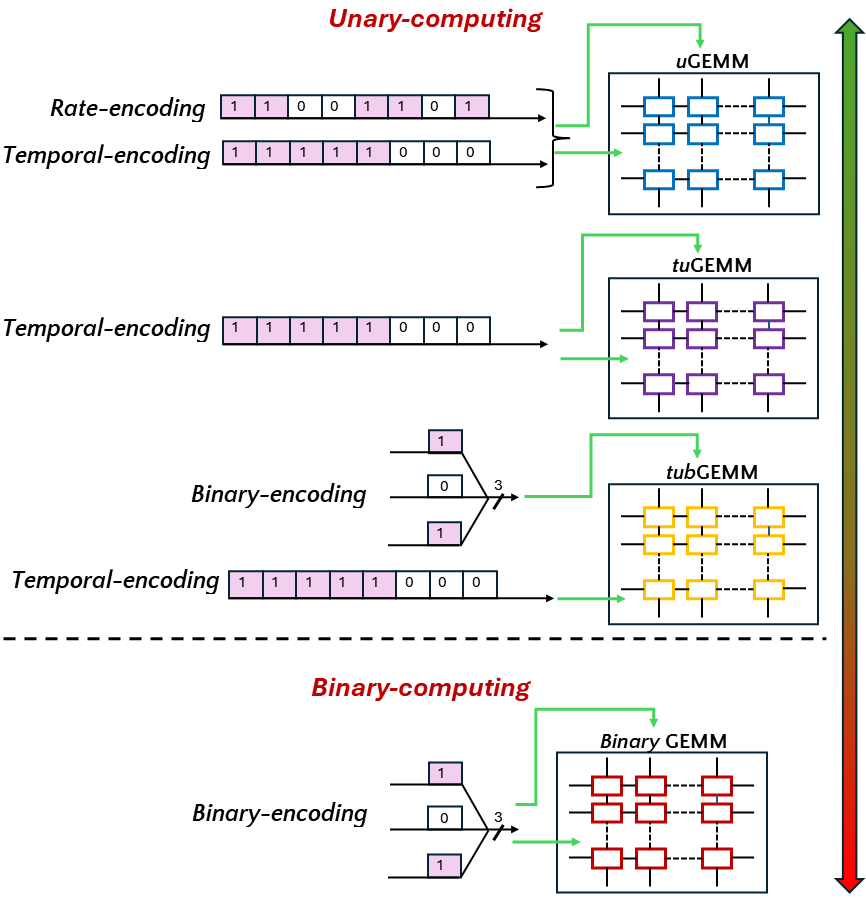} 
\caption{Four GEMM designs: uGEMM encodes both inputs as unary (rate/temporal), \textit{tu}GEMM encodes both inputs as temporal-unary, \textit{tub}GEMM encodes one input as temporal-unary and other binary, and the conventional binary GEMM. The vertical ordering of the four designs reflects a spectrum from \textit{fully-unary} design (green) to a \textit{fully-binary} design (red).}
\label{fig:overview}
\end{figure}


We first review the three unary GEMM designs (Figure \ref{fig:overview}).
%




\subsection{uGEMM}
uGEMM \cite{ugemm_paper} adopts a \textit{unified}-unary GEMM architecture, supporting both rate-coding and temporal-coding. It introduces \textit{unified}-unary computing units for multiplication (uMULT), scaled addition (uSADD), and non-scaled addition (uNSADD). To mitigate long latency and high energy consumption, it incorporates \textit{early termination}, reporting a 98\% increase in energy efficiency and a minimal loss of less than 0.5\% from the ideal output. While being the first to support fully streaming execution, a non-streaming variant is evaluated for a fair comparison with other designs, using binary inputs converted to unary-coded inputs.
uGEMM requires $2^{w}$ cycles for a GEMM computation, with $w$ as bitwidth. Parallel adder trees are used to accumulate all partial sum bitstreams.

\subsection{tuGEMM}

\textit{tu}GEMM \cite{tugemm} is the pioneering counter-based GEMM architecture supporting deterministic computing with fully temporal-encoding. Two design variants, \textit{serial} and \textit{parallel}, offer different area-latency tradeoffs. \textit{tu}GEMM achieves significantly lower area and power consumption but at the cost of quadratically worse latency relative to uGEMM. Authors in \cite{tugemm} present post-synthesis results for different low bit-precisions and matrix sizes but comparison with uGEMM is only performed for 8-bit 16x16 GEMM. Latency for \textit{tu}GEMM scales exponentially with bitwidth due to operation on nested temporal-coded bitstreams, leading to a worst-case latency of $N*{(2^{w-1})}^2$ cycles for a GEMM compute, with $N$ as the common dimension of input matrices and $w$ as the bitwidth.

\subsection{tubGEMM}
\textit{tub}GEMM \cite{vellaisamy2023tubgemm} utilizes hybrid temporal-unary and binary (tub) encoding, performing exact (deterministic) computation using sequential multipliers and accumulators. Compared to uGEMM, it achieves substantial reductions in area (89\%), power (87\%), and energy (50\%) for 8-bit 16x16 GEMM in 45nm CMOS. Notably, authors in \cite{vellaisamy2023tubgemm} also evaluate \textit{tub}GEMM across varying bitwidths and matrix sizes on TSMC N5 process node. Relative to \textit{tu}GEMM, \textit{tub}GEMM reduces worst-case latency to $N*{(2^{w-2})}$ cycles, optimizing its temporal encoding with a novel $2$\textit{-unary} scheme that halves the latency.

\section{Evaluation Methodology}

This section details the evaluation methodology for post-synthesis Power-Performance-Area (PPA), ensuring fair comparison among all GEMM designs. Additionally, the setup for weight sparsity analysis profiled for DL models is described.

\subsection{Setup for PPA Analysis}
All GEMM designs are synthesized using Nangate45 open-source library with Synopsys Design Compiler, operating at a clock frequency of 400 MHz. Synthesis is performed exhaustively for 2-, 4-, and 8-bits across two matrix sizes: 16x16 and 32x32. The chosen bitwidths and matrix sizes are amenable for edge AI inference in modern mobile system-on-chips (SoCs). All GEMM designs implement non-scaled bipolar compute \cite{ugemm_paper} and utilize the outer product dataflow as in \cite{tugemm, vellaisamy2023tubgemm}. The binary-based GEMM (`bGEMM') serves as our baseline benchmark. bGEMM is synthesized using DesignWare multipliers and adders, and incurs $N$ cycles for a single GEMM compute with outer product dataflow. 
Along with area and power results, we further derive energy and area-delay product (ADP) metrics based on the GEMM latencies. Note that ADP inherently captures and normalizes the spatio-temporal trade-offs in the unary GEMM designs, such as in serial (considered here) vs. parallel \textit{tu}GEMM (omitted for brevity).
Worst-case (WC) latency is calculated by multiplying compute cycles (Sec. \ref{sec:gemms}) with the clock period (2.5 ns).

%
%
\subsection{Setup for Weight Sparsity Analysis}
Two types of weight sparsities are considered: 1) Word sparsity signifies zero values (percentage of weights that have zero magnitude); 2) Bit sparsity denotes small magnitude values (percentage of `0' bits in the temporal-unary bitstream). In the extreme case when all bits are `0's, bit sparsity subsumes word sparsity. Higher bit sparsity (\textit{b\_spa}) implies a large number of `0' bits (i.e., small number of `1' bits) in the temporal-unary bitstream, leading to lower dynamic latency (and energy consumption) for \textit{tu}GEMM and \textit{tub}GEMM:
\begin{equation}
\label{dyn-latency}
    \text{Dynamic Latency} = \text{WC Latency} * (1 - b\_spa)
\end{equation}

Eight pretrained quantized INT8 CNNs are imported from Torchvision for sparsity profiling: 1) MobileNetV2, 2) MobileNetV3, 3) InceptionV3, 4) ShuffleNetV2, 5) GoogleNet, 6) ResNet18, 7) ResNet50, and 8) ResNeXt101. These models are widely used in leading DL literature.
Similar to the methodology in \cite{vellaisamy2023tubgemm}, maximum values within each feature map are tracked, and the corresponding counts are averaged across convolution and fully connected layers to derive bit sparsity.

Additionally, a pretrained quantized INT32 LLaMA2-70B model from Huggingface is profiled.
Due to constraints on accessing low-precision LLaMA2 model weights, a pragmatic approach is adopted by considering the most significant bits (MSBs) for sparsity profiling, as used in prior works~\cite{choi2019accurate, truncation_quantization_paper2} without impacting the distribution and sparsity significantly. The weights are extracted from two fully connected (FC) layers in the attention layer and a 2-layer feed-forward network (FFN) after the attention layer. We also profile the query (Q) and key (K) tokens in the attention layer, using small representative inputs.
Average maximum value per 32x32 block is considered, as largest value bottlenecks GEMM compute.

\section{Experimental Results}

\begin{table}[t]
\caption{45nm post-synthesis area (in $\mu$m$^2$) for the four GEMM designs with varying bit-widths and matrix sizes.}
\label{tab:area}
\centering
\resizebox{0.95\columnwidth}{!}{%
\begin{tabular}{|cc|c|c|c|c|}
\hline
\multicolumn{2}{|c|}{\textbf{Configuration}}                    & \textbf{uGEMM} & \textbf{\textit{tu}GEMM} & \textbf{\textit{tub}GEMM} & \textbf{bGEMM} \\ \hline 
\multicolumn{1}{|c|}{\multirow{2}{*}{\textbf{2-bit}}} & \textbf{16x16}   &  \textbf{\textcolor{red}{99,445.7}}     & \textbf{\textcolor{DarkGreen}{13,436.4}}    &  19,112.6       & 16,739.1        \\ \cline{2-6} 
\multicolumn{1}{|c|}{}                       & \textbf{32x32}   & \textbf{\textcolor{red}{791,794.4}}     & \textbf{\textcolor{DarkGreen}{52,272.4}}       &  76,375.5      & 67,201.7      \\ \cline{2-6} 
\hline
\multicolumn{1}{|c|}{\multirow{2}{*}{\textbf{4-bit}}} & \textbf{16x16}   &  \textbf{\textcolor{red}{203,920.7}}     & \textbf{\textcolor{DarkGreen}{29,061.0}}       &  38,912.6       & 44,925.8      \\ \cline{2-6} 
\multicolumn{1}{|c|}{}                       & \textbf{32x32}   &  \textbf{\textcolor{red}{1,799,961.0}}    & \textbf{\textcolor{DarkGreen}{117,261.3}}    & 151,933.6        & 180,458.6    \\ \cline{2-6} 
\hline
\multicolumn{1}{|c|}{\multirow{2}{*}{\textbf{8-bit}}} & \textbf{16x16}   &  \textbf{\textcolor{red}{445,396.2}}   & \textbf{\textcolor{DarkGreen}{61,064.0}}        &   99,916.8     & 132,786.9      \\ \cline{2-6} 
\multicolumn{1}{|c|}{}                       & \textbf{32x32}   & \textbf{\textcolor{red}{3,689,829.0}}    & \textbf{\textcolor{DarkGreen}{235,470.9}}       &  338,692.7       & 560,778.5      \\ \cline{2-6} 
\hline
\end{tabular}%
}

\end{table}

This section presents four types of evaluation for the GEMM designs (uGEMM, \textit{tu}GEMM, \textit{tub}GEMM, bGEMM) across 8, 4, 2-bit precisions and 16x16, 32x32 matrix sizes: 1) 45nm CMOS post-synthesis area-power results, 2) Energy results derived from worst-case latencies, 3) Area-Delay Product (ADP) results derived from worst-case latencies, and 4) Workload-dependent sparsity and corresponding latency-energy results. Further, PPA results are also shown for larger 64x64 and 128x128 matrix sizes for 4-bit precision.

\begin{table}[t]
\caption{45nm post-synthesis power (in mW) for the four GEMM designs with varying bit-widths and matrix sizes.}
\label{tab:power}
\centering
\resizebox{0.95\columnwidth}{!}{%
\begin{tabular}{|cc|c|c|c|c|}
\hline
\multicolumn{2}{|c|}{\textbf{Configuration}}                    & \textbf{uGEMM} & \textbf{\textit{tu}GEMM} & \textbf{\textit{tub}GEMM} & \textbf{bGEMM} \\ \hline 
\multicolumn{1}{|c|}{\multirow{2}{*}{\textbf{2-bit}}} & \textbf{16x16}   &   \textbf{\textcolor{red}{42.2}}    & \textbf{\textcolor{DarkGreen}{4.9}}       &  5.0       &  7.7     \\ \cline{2-6} 
\multicolumn{1}{|c|}{}                       & \textbf{32x32}   &  \textbf{\textcolor{red}{323.8}}    &  \textbf{\textcolor{DarkGreen}{18.3}}      &  19.8       & 30.9      \\ \cline{2-6} 
\hline
\multicolumn{1}{|c|}{\multirow{2}{*}{\textbf{4-bit}}} & \textbf{16x16}   &  \textbf{\textcolor{red}{64.1}}    & \textbf{\textcolor{DarkGreen}{9.2} }      &  9.9       &  22.4     \\ \cline{2-6} 
\multicolumn{1}{|c|}{}                       & \textbf{32x32}   &  \textbf{\textcolor{red}{513.6}}    &  \textbf{\textcolor{DarkGreen}{37.2} }     &  39.1       & 88.3      \\ \cline{2-6} 
\hline
\multicolumn{1}{|c|}{\multirow{2}{*}{\textbf{8-bit}}} & \textbf{16x16}   &  \textbf{\textcolor{red}{100.8}}     & \textbf{\textcolor{DarkGreen}{19.7} }      &  26.1       &  72.8     \\ \cline{2-6} 
\multicolumn{1}{|c|}{}                       & \textbf{32x32}   &   \textbf{\textcolor{red}{784.4}}    &  \textbf{\textcolor{DarkGreen}{74.7} }     &  90.9       &  321.3     \\ \cline{2-6} 
\hline
\end{tabular}%
}
\end{table}

\subsection{Area-Power Evaluation}
\label{subsec:area-power}

\begin{figure}[t]
\centering
\includegraphics[width=0.83\columnwidth]{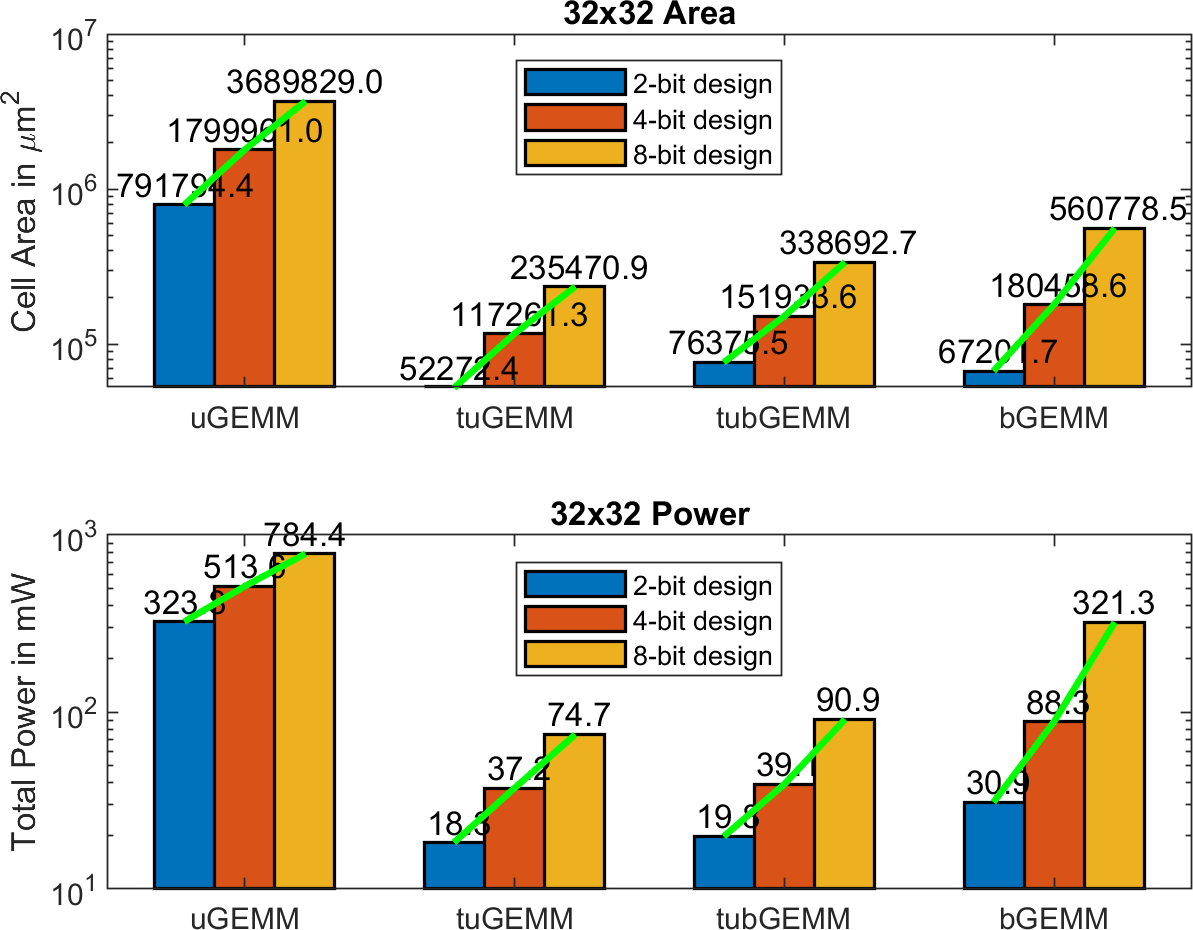} 
\caption{45nm post-synthesis area and power scaling across 2, 4, and 8-bits for 32x32 matrix size. Y-axis is in log scale.}
\label{fig:area_bar}
\end{figure}

Tables~\ref{tab:area} and \ref{tab:power} illustrate the 45nm post-synthesis cell area and total power, and are pivotal for gauging trade-offs in GEMM hardware efficiency for AI inference. The lowest (best) and highest (worst) values are marked in green and red, respectively. \textit{tu}GEMM outperforms all other designs in area-power efficiency across all configurations owing to its simplistic counter-based streamlined architecture without the need for any huge adder trees. As it relies predominantly on counters for the temporal accumulation of vector-vector products, it incurs very high latency leading to highest energy consumption among all designs (Table \ref{tab:energy} as will be explained in Sec. \ref{subsec:wc-energy}).
\textit{tub}GEMM is the next optimal design in area-power efficiency across 4-bits and 8-bits.
bGEMM outperforms \textit{tub}GEMM in area at 2-bits but scales considerably worse with increasing bitwidth. uGEMM, while versatile in supporting both rate-coding and temporal-coding, exhibits lower area-power efficiency across all designs due to its unified unary approach.
Figure \ref{fig:area_bar} plots these results for 32x32 GEMM.
More specifically, uGEMM, \textit{tub}GEMM and bGEMM are approximately 7x, 1.5x and 2x worse than \textit{tu}GEMM on average for 16x16 GEMM area. uGEMM scales poorly with matrix size resulting in an increased gap of 15x for 32x32 GEMM, while the gap remains consistent for \textit{tub}GEMM and bGEMM, implying \textit{tu}GEMM, \textit{tub}GEMM and bGEMM scale similarly with matrix sizes.
In power, \textit{tu}GEMM and \textit{tub}GEMM are close and consistently outperform uGEMM and bGEMM, with uGEMM consuming the most power (about 10x for 8-bit 32x32 compared to \textit{tu}GEMM). The power efficiency of \textit{tu}GEMM and \textit{tub}GEMM is superior mainly because temporal-unary encoding results in only two signal transitions due to consecutive ones followed by zeros, in contrast to rate-unary and binary encoding with multiple signal transitions.

In terms of bitwidth scaling, all designs scale linearly on log scale (green trendlines in Figure \ref{fig:area_bar}). However, the slopes are different for each of the designs (lower slope indicates better scaling). Specifically, for area, \textit{tu}GEMM and \textit{tub}GEMM scale best (2.12 slope), closely followed by uGEMM (2.16) and finally bGEMM (2.90). Similarly, the four designs incur slopes of 2.02, 2.15, 1.56 and 3.25 respectively for power, indicating best scaling for uGEMM. uGEMM's superior bitwidth scaling is due to its simpler stochastic single-gate multiplier. In contrast, with increasing matrix sizes, the adder trees in uGEMM become significantly denser resulting in poor scaling.


\textbf{Key Takeaway:} \textit{tu}GEMM has the best area-power efficiency, closely followed by \textit{tub}GEMM and bGEMM. uGEMM is almost 10x worse. \textit{tu}GEMM and \textit{tub}GEMM scale well with bitwidths and matrix sizes. uGEMM scales well with bitwidths but poorly with matrix sizes, and vice versa for bGEMM.


\subsection{Energy Evaluation with Worst-Case Latency}
\label{subsec:wc-energy}

Table \ref{tab:energy} shows energy consumption based on worst-case latency for each GEMM design.
Despite its optimal area-power efficiency, \textit{tu}GEMM incurs the highest energy due to its nested counter-based temporal accumulation, resulting in significant latency for 8-bits and above. bGEMM is the most energy-efficient for 4- and 8-bits owing to just $N$ cycles per GEMM independent of bitwidth, whereas \textit{tub}GEMM outperforms bGEMM for 2-bits. \textit{tub}GEMM is reasonably close (1.8x) to bGEMM's energy consumption at 4 bits, with uGEMM trailing behind (2.9x). With these findings, 4-bit precision emerges as a focal point for detailed analysis. Further post-synthesis results for larger Google EdgeTPU (64x64) and CloudTPUv3 (128x128) matrix sizes are detailed in Table \ref{tab:tpu} for 4-bits. It shows \textit{tu}GEMM as most efficient in area and power but least in energy, as expected.
Notably, \textit{tub}GEMM consumes just 1.2x more energy than bGEMM for EdgeTPU (64x64) but outperforms bGEMM at CloudTPUv3 (128x128) array size, resulting in 12\% more energy efficiency, even with worst-case latency. This suggests optimal scalability for \textit{tub}GEMM targeting DLAs with large (beyond 64x64)  4-bit and 2-bit processing element (PE) arrays.

\textbf{Key Takeaway:} \textit{tub}GEMM is most energy-efficient at 2-bits, and is comparable with bGEMM at 4-bits. For large PE arrays, \textit{tub}GEMM emerges as a better candidate than bGEMM, owing to its high scalability of energy efficiency with matrix sizes, even with worst-case latency.

\begin{table}[t]
\caption{45nm post-synthesis energy (in nJ) for worst-case GEMM latencies with varying bit-widths and matrix sizes.}
\label{tab:energy}
\centering
\resizebox{0.9\columnwidth}{!}{%
\begin{tabular}{|cc|c|c|c|c|}
\hline
\multicolumn{2}{|c|}{\textbf{Configuration}}                    & \textbf{uGEMM} & \textbf{\textit{tu}GEMM} & \textbf{\textit{tub}GEMM} & \textbf{bGEMM} \\ \hline 
\multicolumn{1}{|c|}{\multirow{2}{*}{\textbf{2-bit}}} & \textbf{16x16}   &   0.42    & \textbf{\textcolor{red}{0.78} }      &  \textbf{\textcolor{DarkGreen}{0.20}}       &  0.31     \\ \cline{2-6} 
\multicolumn{1}{|c|}{}                       & \textbf{32x32}   &  3.24     &  \textbf{\textcolor{red}{5.86}}      &  \textbf{\textcolor{DarkGreen}{1.58}}      & 2.47      \\ \cline{2-6} 
\hline
\multicolumn{1}{|c|}{\multirow{2}{*}{\textbf{4-bit}}} & \textbf{16x16}   &    2.56   & \textbf{\textcolor{red}{23.55}}       &  1.58       &  \textbf{\textcolor{DarkGreen}{0.90} }    \\ \cline{2-6} 
\multicolumn{1}{|c|}{}                       & \textbf{32x32}   &  20.54     &  \textbf{\textcolor{red}{190.46}}      &  12.51       & \textbf{\textcolor{DarkGreen}{7.06}}      \\ \cline{2-6}
\hline
\multicolumn{1}{|c|}{\multirow{2}{*}{\textbf{8-bit}}} & \textbf{16x16}   &  64.51     & \textbf{\textcolor{red}{12,910.59}}    &  66.82       &  \textbf{\textcolor{DarkGreen}{2.91}}     \\ \cline{2-6} 
\multicolumn{1}{|c|}{}                       & \textbf{32x32}   &   502.02    & \textbf{ \textcolor{red}{97,910.78} }   &  465.41       &  \textbf{\textcolor{DarkGreen}{25.70}}     \\ \cline{2-6}
\hline
\end{tabular}%
}
\end{table}

\begin{table}[t]
\caption{45nm post-synthesis area, power, and energy (with worst-case latency) for EdgeTPU (64x64) and Cloud TPUv3 (128x128) GEMM sizes for 4-bit precision.}
\label{tab:tpu}
\centering
\resizebox{0.95\columnwidth}{!}{%
\begin{tabular}{|cc|c|c|c|c|}
\hline
\multicolumn{2}{|c|}{\textbf{Configuration}}                    & \textbf{uGEMM} & \textbf{\textit{tu}GEMM} & \textbf{\textit{tub}GEMM} & \textbf{bGEMM} \\ \hline 
\multicolumn{1}{|c|}{{\textbf{4-bit Area}}} & \textbf{64x64}   &   \textbf{\textcolor{red}{15.89}}    &  \textbf{\textcolor{DarkGreen}{0.46}}   & 0.59     &   1.09    \\ \cline{2-6} 
\multicolumn{1}{|c|}{(mm$^2$)}                         & \textbf{128x128}   & \textbf{\textcolor{red}{140.24}}  &  \textbf{\textcolor{DarkGreen}{1.83}} &   2.41     &  6.64 \\  \cline{2-6}
\hline
\hline 
\multicolumn{1}{|c|}{\multirow{1}{*}{\textbf{4-bit Power}}} & \textbf{64x64}   &  \textbf{\textcolor{red}{4,115.21}} &  \textbf{\textcolor{DarkGreen}{145.52}}   &   154.42   &   496.77    \\ \cline{2-6} 
\multicolumn{1}{|c|}{(mW)}                       & \textbf{128x128}   & \textbf{\textcolor{red}{32,973.04}} & \textbf{\textcolor{DarkGreen}{579.28}} &   620.92     &  2,794.80   \\  \cline{2-6}
\hline
\hline 
\multicolumn{1}{|c|}{\multirow{1}{*}{\textbf{4-bit Energy}}} & \textbf{64x64}   &  164.61 &  \textbf{\textcolor{red}{1,490.12}}   &  98.83  &  \textbf{\textcolor{DarkGreen}{79.48}}     \\ \cline{2-6} 
\multicolumn{1}{|c|}{(nJ)}                       & \textbf{128x128}  & 1,318.92 & \textbf{\textcolor{red}{11,863.65}}  &      \textbf{\textcolor{DarkGreen}{794.78}}  &  894.34   \\  \cline{2-6}
\hline
\hline
\multicolumn{1}{|c|}{{\textbf{4-bit ADP}}} & \textbf{64x64}   &  635.6    &   \textcolor{Red}{\textbf{4,710.4}}  &  377.6    &  \textcolor{DarkGreen}{\textbf{174.4}}    \\ \cline{2-6} 
\multicolumn{1}{|c|}{(mm$^2$-ns)}                         & \textbf{128x128}   & 5,609.6 & \textcolor{Red}{\textbf{37,478.4}}  & 3,084.8    & \textcolor{DarkGreen}{\textbf{2,124.8}}   \\  \cline{2-6}
\hline


\end{tabular}%
}
\end{table}

\subsection{Area-Delay Product Evaluation with Worst-Case Latency}
Table \ref{tab:tpu} also shows Area-Delay Product (ADP) values for 64x64 and 128x128 matrix sizes. ADP is a useful metric to analyse the spatio-temporal trade-offs of GEMM designs.
Table \ref{tab:tpu} shows that bGEMM has the lowest (best) ADP due to its minimal latency, closely followed by \textit{tub}GEMM with 2.2x and 1.5x higher ADP for 64x64 and 128x128 arrays, respectively (this gap is reduced with increasing matrix sizes). uGEMM's ADP is \url{~}3x higher than bGEMM on average and \textit{tu}GEMM's ADP is \url{~}20x higher (infeasibly large).

\textbf{Key Takeaway:} Despite unary designs having better area than bGEMM, the substantial latency increase results in worse ADP, indicating potential room for area-latency improvement.




\begin{table}[t]
    \caption{Profiled weight sparsities of CNNs and LLM. Word sparsity denotes percentage of zero weights; bit sparsity denotes percentage of zero bits within temporal-unary weights.}
    \label{tab:sparsity}
    \centering
    \begin{adjustbox}{max width=0.95\linewidth}
    \begin{tabular}{c|c|c}
        \toprule
        \textbf{CNN} & \textbf{Word (\%) 8 bits} & \textbf{Bit (\%) 8 bits}  \\
        \midrule
        \midrule
        MobileNetV2 & 2.25 & 44.66 \\
        \midrule
        MobileNetV3 & 9.52 & 38.59 \\
        \midrule
        GoogleNet & 1.91 & 45.91 \\
        \midrule
        InceptionV3 & 1.99 & 45.61 \\
        \midrule
        ShuffleNetV3 & 1.43 & 47.18 \\
        \midrule
        ResNet18 & 2.04 & 45.3 \\
        \midrule
        ResNet50 & 2.45 & 46.24 \\
        \midrule
        ResNeXt101 & 2.64 & 44.23 \\
        \midrule 
        \midrule
        \textbf{LLaMA2-70B Layer (LLM)} & \textbf{Word (\%) 2/4/8 bits} & \textbf{Bit (\%) 2/4/8 bits} \\
        \midrule
        \midrule
        Attention FC layer weight & 20.7 / 2.85 / 0.0613 & 50.00 / 12.50 / 0.82 \\
        \midrule
        FFN layer weight & 20.8 / 3.02 / 0.0524 & 50.00 / 12.5 / 0.80 \\
        \midrule
        Self attention $Q$ & 82.8 / 35.0 / 2.71 & 0.56 / 8.89 / 28.84 \\
        \midrule
        Self attention $K$ & 85.1 / 37.4 / 2.94 & 8.19 / 8.58 / 32.52 \\
        \bottomrule
    \end{tabular}
    \end{adjustbox}
\end{table}


\begin{figure}[t]
\centering
\includegraphics[width=0.9\columnwidth, height=5.2cm]{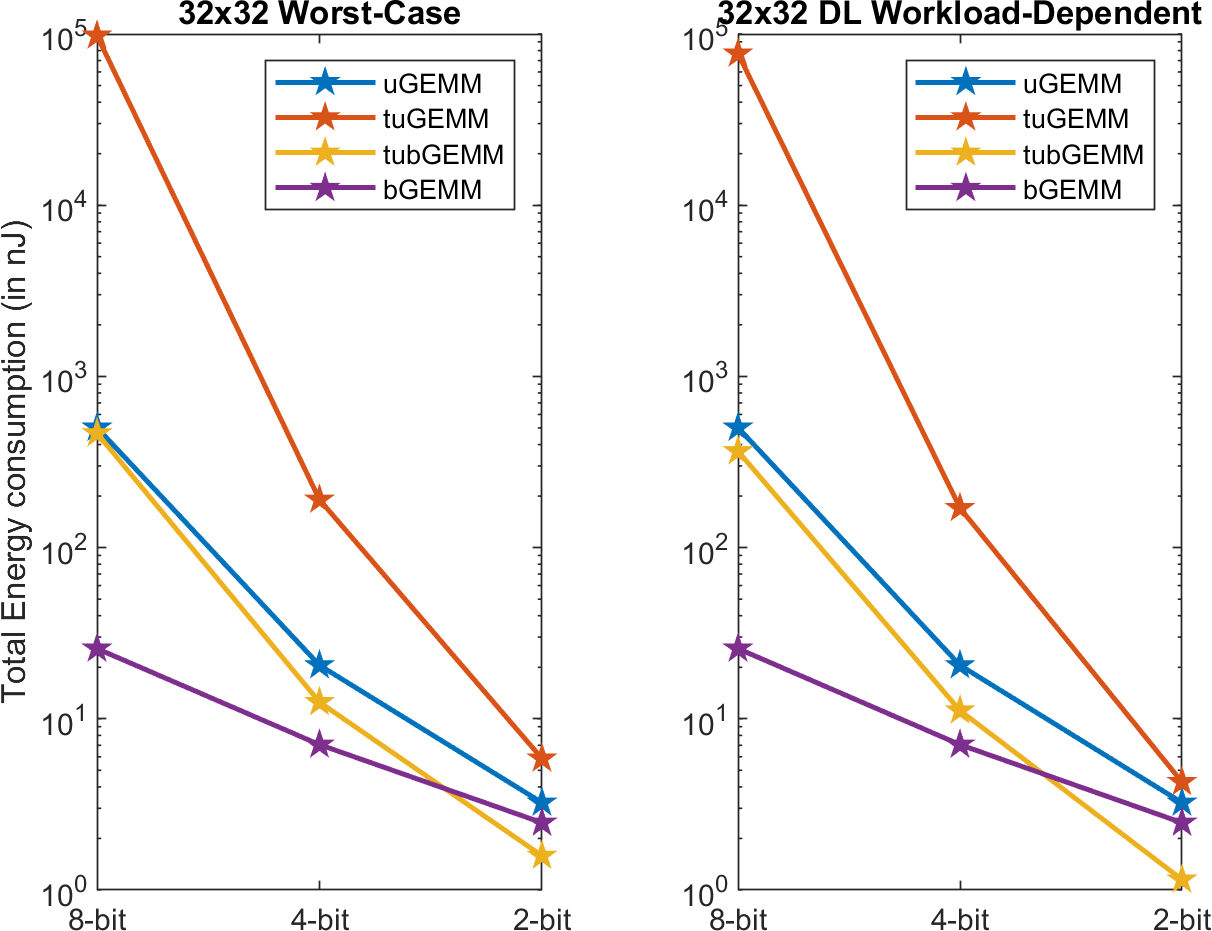} 
\caption{Energy consumption for 32x32 GEMM across 8-, 4- and 2-bits. Y-axis is in log scale. Compared to left worst-case plot, note for \textit{tubGEMM}, increased energy efficiency at 2 bits, earlier cross-over point with bGEMM, and larger energy gap to uGEMM at 8 bits, with sparsity (right plot).}
\label{fig:energy_bar}
\end{figure}

\subsection{Sparsity-Driven Latency-Energy Evaluation}
\label{sec:sparsity}

The weight sparsity profiling results are summarized in Table~\ref{tab:sparsity}. Across most CNN models, there is a consistent word sparsity of approximately 2\%, with MobileNetV3 being the outlier with 9.5\% of its weights as zeros. For the LLaMA2-70B LLM model, the 8-bit self-attention tokens exhibit comparable weight sparsity, averaging around 2.8\% (equivalent to approximately 2B zero weights out of the total 70B weights). 4-bit (36\%) and 2-bit (84\%) word sparsities are significantly higher. In contrast, the attention FC and FFN layers involve much lower word sparsities (negligible for 8-bits).

Bit sparsity subsumes word sparsity and directly translates to latency and energy improvements for \textit{tu}GEMM and \textit{tub}GEMM.
CNNs display significantly better bit sparsity (\url{~}43\%) compared to LLM (negligible for 8-bit FC/FFN layers and \url{~}30\% for tokens). 4- and 2-bit values show varying sparsities for LLM. Plugging in the bit sparsity values from Table \ref{tab:sparsity} (in fractional form) into Equation \ref{dyn-latency}, DL workload-dependent energy values are derived for 8-, 4-, and 2-bits for 32x32 \textit{tu}GEMM and \textit{tub}GEMM. Note, only the two temporal-unary designs can leverage bit sparsity. These values are plotted to the right in Figure \ref{fig:energy_bar} with worst-case energy values to the left. Three notable improvements for \textit{tub}GEMM upon leveraging sparsity are: 1) Enhanced 2-bit energy efficiency, further increasing the gap with bGEMM. 2) Earlier cross-over point with bGEMM, indicating \textit{tub}GEMM can now outperform or perform on par with bGEMM for 3-bits. 3) More discernable energy gap with uGEMM at 8-bits.

\textbf{Key Takeaway:} 
Overall, \underline{\textit{tub}GEMM} stands out as the best design for low-precision AI inference (4 and 2 bits) due to its high area-power-energy efficiency, further enhanced through bit sparsity in DL workloads. \textit{tu}GEMM's low hardware complexity makes it reasonable for applications (especially 2 bits) where area and power are highly constrained but high latency can be tolerated. uGEMM is suitable where the compute infrastructure demands rate-unary inputs, particularly for small matrix sizes. Finally, bGEMM is desirable for low-latency compute, especially for 8-bits and above.


\section{Potential for Unary-based AI Compute}

Our findings suggest  the following areas for further exploring of unary based designs for energy-efficient  AI inference:

\subsubsection{Temporal-Unary Compute}
While rate-unary methods in stochastic computing have been widely studied, temporal-unary computing remains under-explored. Temporal-unary designs like \textit{tu}GEMM and \textit{tub}GEMM, which use standard digital circuits, show promise for AI applications. Stochastic computing can lead to accuracy loss, as seen when a 96.08\% accurate INT8 quantized MLP model drops to 94.7\% with uGEMM. Temporal-unary, however, offers deterministic computation without this accuracy degradation. The \textit{n}-unary encoding from \cite{vellaisamy2023tubgemm} presents opportunities for further latency and energy efficiency improvements, with possible schemes to offset any trade-offs. With the \textit{tub}GEMM design displaying PPA results close to conventional binary GEMM design, further design optimization can lead to it outperforming across all configurations, especially for weights of 2-4 bits.

\subsubsection{Leveraging Sparsity}
The sparsity results in Table~\ref{tab:sparsity} and analysis in Sec. \ref{sec:sparsity} highlight opportunities for bit sparsity exploitation. Given the growing importance of pruning and SparseML libraries in edge AI, naturally exploiting sparsity, as seen in \textit{tu}GEMM and \textit{tub}GEMM, offers substantial prospects for improving latency and energy efficiency. 


\subsubsection{Low-Precision Compute}
Given the ongoing advancements in quantization techniques \cite{wang20188, choi2019accurate, ma2024era, truncation_quantization_paper2}, AI inference is progressively embracing ultra-low bit-precisions. This trend affords unary designs the opportunity for exponential reductions in compute latency, rendering them a viable alternative to binary-based processing elements for 4 bits and below.

\bibliographystyle{IEEEtran}
\bibliography{ref}

\end{document}